\documentclass[journal]{IEEEtran}

\usepackage{graphicx}
\usepackage{epstopdf} 
\usepackage{xcolor}
\usepackage{amsmath}
\usepackage{amsfonts}
\usepackage{fixltx2e}
\usepackage{hyperref}
\usepackage{braket}
\usepackage{subcaption}
\usepackage{multirow}

\usepackage{ifpdf}
\usepackage{cite}

\begin{document}

\title{Quantum Stream Learning}


\author{Yongcheng Ding,  Xi Chen,  Rafael Magdalena-Benedicto and~Jos{\'e} D.  Mart{\'i}n-Guerrero, {\em Member, IEEE}

\thanks{Yongcheng Ding and Xi Chen are with Department of Physical Chemistry, University of the Basque Country UPV/EHU, Spain.}
\thanks{Yongcheng Ding is also with International Center of Quantum Artificial Intelligence for Science and Technology (QuArtist)  and Department of Physics, Shanghai University, 200444 Shanghai, China
	 and ProQuam Co, Ltd., Shanghai.}
\thanks{Jos{\'e} D.  Mart{\'i}n-Guerrero and Rafael Magdalena-Benedicto are with IDAL,  Departament of Electronic Engineering, ETSE-UV,
       Universitat de Val\`{e}ncia (UV), Spain.}
        }

\markboth{IEEE Transactions on Neural Networks and Learning Systems,~Vol.~XX, No.~YY, 2021}%
{Shell \MakeLowercase{\textit{et al.}}: Bare Demo of IEEEtran.cls for IEEE Journals}

\maketitle

\begin{abstract}
The exotic nature of quantum mechanics makes machine learning (ML) be different in the quantum realm compared to classical applications.  ML can be used for knowledge discovery using information continuously extracted from a quantum system in a broad range of tasks. The model receives streaming quantum information for learning and decision-making, resulting in instant feedback on the quantum system. As a stream learning approach, we present a deep reinforcement learning on streaming data from a continuously measured qubit at the presence of detuning, dephasing, and relaxation. We also investigate how the agent adapts to another quantum noise pattern by transfer learning. Stream learning provides a better understanding of closed-loop quantum control, which may pave the way for advanced quantum technologies.
\end{abstract}


\IEEEpeerreviewmaketitle

\section{Introduction}
\label{sec:Introduction}

\IEEEPARstart{Q}{uantum} computation and quantum information~\cite{nielsen2002quantum} are no longer promising research fields but current realities whose applicability is and will be increasing in the next decade, in particular, that related to computational intelligence~\cite{manju2014,Nguyen2020}.  Quantum computation is driven by the so-called quantum bits (qubits), a quantum generalization of the classical bit. The two basic states of a qubit are $\ket{0}$ and $\ket{1}$,  corresponding with the states zero and one, respectively, of a classical bit. A qubit $\ket{\Psi}$ generalizes its classical counterpart because it allows states formed by the superposition of $\ket{0}$ and $\ket{1}$, namely, $\ket{\Psi}=\alpha \ket{0} + \beta \ket{1}$, where $\alpha$ and $\beta$ are complex coefficients. The measurement of a qubit in superposition state involves that it will collapse to one of its basic states, but there is no way to determine in advance which one; the unique available information is that the probability of $\ket{0}$ is $|\alpha|^2$ and the probability of $\ket{1}$ is $|\beta|^2$, hence, $|\alpha|^2 + |\beta|^2=1$. The main operation when dealing with qubits is the unitary transformation $U$. When $U$ is applied to a superposition state, the result is another superposition state that superposes all basis vectors. This is an appealing characteristic of unitary transformations, which is called quantum parallelism because it can be employed to evaluate the different values of a function f(x) for a given input x at the same time. The unitary transformation $U(t,t_0)$ evolves a qubit state $\ket{\Psi(t_0)}$ to $\ket{\Psi(t)}=U(t,t_0)\ket{\Psi(t_0)}=\mathcal{T}\exp[-(i/\hbar)\int_{t_0}^tH(t')dt']\ket{\Psi(t_0)}$, where $H(t')$ is its Hamiltonian. Consequently, quantum control arises as the most critical problem in realizing quantum computation. It aims at designing a time-dependent Hamiltonian $H(t)$, which drives the qubit to its target by a unitary transform. It can be indeed trivial, e.g., one may have the quantum NOT gate by a resonant pulse $H=\hbar\Omega\sigma_x/2$, flipping a qubit within operation time $T=\hbar\pi/2\Omega$. However, the solution is far from optimal since it is not robust, i.e., any slight systematic error $T\rightarrow T+\delta T$ or equivalently $\Omega\rightarrow\Omega+\delta\Omega$ will lead to fidelity loss. Meanwhile, qubits can never be perfectly isolated from the external environment, where quantum noises induce decoherence. In this way, optimal quantum control enables the implementations of high-fidelity and high-robustness gate operations, which will be the milestones in fault-tolerant universal quantum computation~\cite{shor1996fault,preskill1998fault,gottesman2010introduction}.

Physicists suggest many protocols for such goals, including adiabatic quantum evolution~\cite{RevModPhys.79.53}, composite pulses~\cite{PhysRevA.70.052318,PhysRevLett.106.233001,rong2015experimental},  pulse-shaping engineering~\cite{PhysRevA.75.062326,PhysRevLett.109.060401,PhysRevLett.111.050404},  shortcuts to adiabaticity~\cite{RevModPhys.91.045001,torrontegui2013shortcuts,PhysRevLett.104.063002}, etc. Particularly, machine learning algorithms (ML) are combined with them for further optimizations~\cite{PhysRevApplied.6.054005,PhysRevLett.123.100501,PhysRevA.103.L040401,ai2021experimentally,PhysRevX.11.031070}. It is also natural to consider applying reinforcement learning (RL) individually for quantum control tasks~\cite{PhysRevX.8.031086,porotti2019coherent,niu2019universal,dalgaard2020global,PhysRevA.97.052333,PhysRevA.99.042327,ostaszewski2019approximation,PhysRevX.11.031070,PhysRevLett.127.190403,chunlin2014,MARTINGUERRERO2022457,app11188589}. In the past few years, deep reinforcement learning (DRL) has solved pulse design for fast and robust quantum state preparation~\cite{henson2018approaching,zhang2019does,haug2020classifying}, gate operation~\cite{an2019deep}, and quantum Szilard engine~\cite{PhysRevA.100.042314}. However, as we highlighted in our previous researches~\cite{PhysRevA.103.L040401,ai2021experimentally}, we are still far from exploiting the power of RL for quantum control because of quantum measurement. RL requires the observation of states for outputting an action, conflicting with quantum mechanics' fundamental feature that the state is destroyed after direct quantum measurement. Most RL models for quantum control are trained in numerical environments instead of real quantum devices for saving resources, followed by fixed pulses after observation and evaluation. These fixed pulses hardly prevail over gradient-based optimization methods. Another approach combines the model with a quantum environment for evaluation, outputting instant action after observing a state. Even though the quantum state is discarded after extraction of quantum information by direct measurement,  the RL model can still realize a quantum control with lack of efficiency if historical actions are stored for repetitive operation of $n(n+1)/2$ steps, retrieving the last destroyed state, where $n$ is the maximum number of timesteps in each episode. 

In this work, we present a stream learning approach to quantum control by employing the RL algorithm for closed-loop quantum control. In this paradigm, qubit's wave functions are no longer destroyed but slightly perturbed after information extraction via weak measurement. The model observes the state, which contains weak values as the partial information of the qubit with less confidence, resulting in an action to evolve the quantum environment to the next timestep. Our scheme reflects the spirit of stream learning once the length of each timestep is sufficiently small, resembling the dynamics of continuous measurement. It also enables transfer learning by adapting the model to the environment during the evaluation while external noises patterns are changing. We reckon that stream learning in the quantum realm enhances the performance of quantum computing and quantum information processing in real-time experiments, accelerating its development from noisy intermediate-scale devices to the next level. 

\begin{figure*}[t]
\centering
\includegraphics[width=1\linewidth]{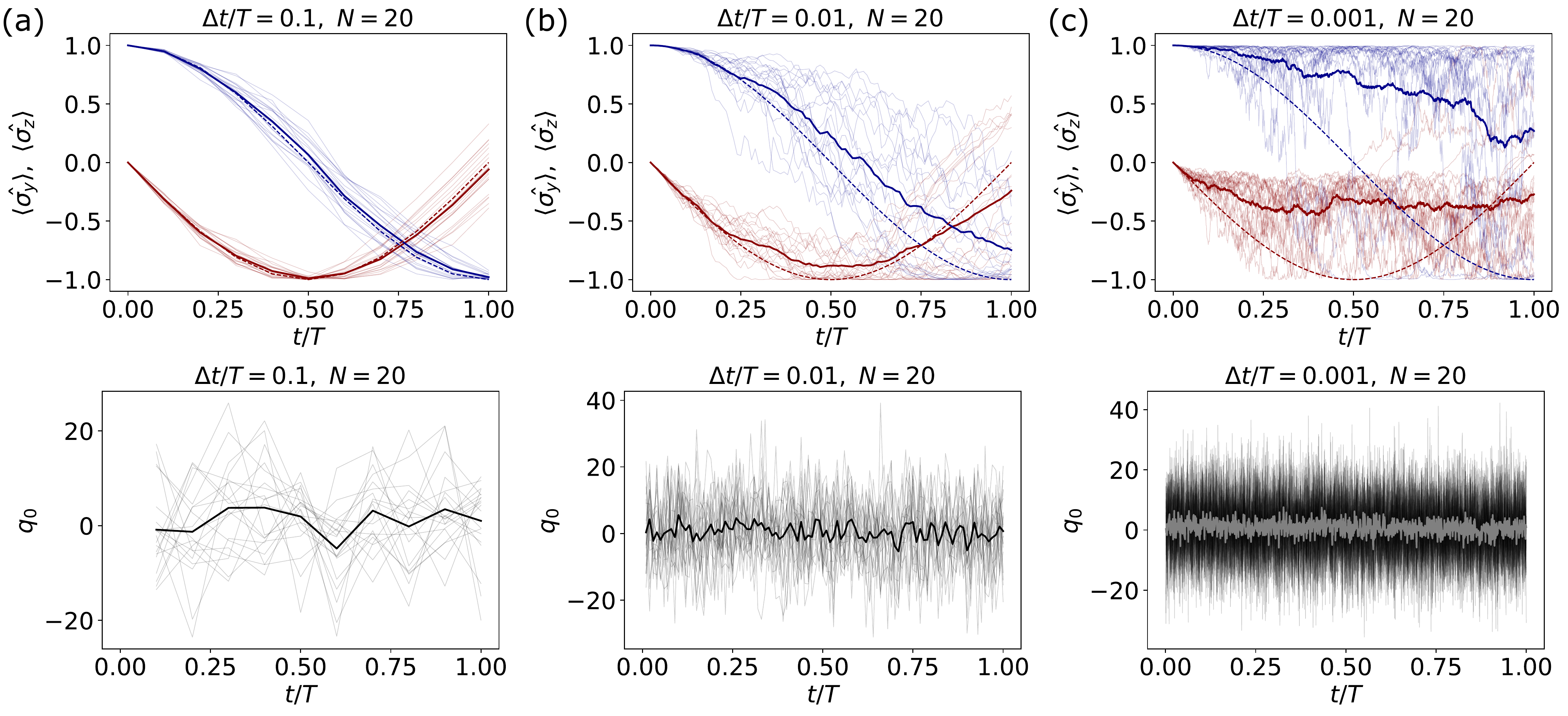}
\caption{The expectations on $Y$ (red) and $Z$ (blue) directions of repetitively measured qubits, when a weak measurement takes place per (a) $\Delta t/T=0.1$, (b) $\Delta t/T=0.01$, and $\Delta t/T=0.001$. The qubit is driven by a resonant $\pi$-pulse, whose measurement-free dynamics is plotted in dashed curves. By averaging over the trajectories in the ensemble of $N=20$ qubits (shaded curves), we have the average trajectories plotted in solid curves. For the closed-loop quantum control, the characteristics of dynamics vary from each other by the scale of the measurement interval. The corresponding weak value feedbacks $q_0$ from measuring the inaccurate Gaussian pointer $\sigma=10$ are also recorded and averaged, showing a low signal-to-noise ratio.  Additionally, we provide the dynamics of a qubit without continuous measurement, being driven by $\pi$-pulse, for comparison purposes. The corresponding expectations are plotted by dashed curves.}
\label{fig:pipulse}
\end{figure*}

\section{Physics models}
\label{sec:models}

\subsection{Open quantum system}
The dynamics of isolated quantum systems are always governed by Schr\"odinger equation $i\hbar\partial_t\ket{\psi(t)}=H(t)\ket{\psi(t)}$, where the operator $H(t)$ is the Hamiltonian of the quantum system, with its expectation be in the unit of energy. It is equivalent to the so-called von Neumann equation $\dot{\rho}(t)=-(i/\hbar)[H(t),\rho(t)]$, where the pure state wave function $\ket{\psi(t)}$ is extended to density matrix $\rho(t)=\ket{\psi(t)}\bra{\psi(t)}$. However, the isolated quantum system never exists in real world. The external environment affects the quantum system by coupling itself to the system, inducing undesired dynamics, such as decoherence. Generally speaking, one can always write down the total Hamiltonian $H_{T}=H_{S}+H_{E}+H_{I}$, including the system Hamiltonian $H_S$, the environment Hamiltonian $H_E$, and the coupling interaction Hamiltonian $H_I$. The new system is assumed to follow the von Neumann equation. One retrieves the information of the original system by tracing out the environmental subsystem $\rho=\text{Tr}_{E}(\rho_T)$, resulting in the Lindblad master equation
\begin{eqnarray}
\label{eq:lindblad}
\dot{\rho}(t)&=&-\frac{i}{\hbar}[H(t),\rho(t)]\nonumber\\
&+&\sum_n\frac{1}{2}\left[2C_n\rho(t)C^\dag_n-\rho(t)C^\dag_nC_n-C^\dag_nC_n\rho(t)\right],
\end{eqnarray}
where $C_n=\sqrt{\gamma_n}A_n$ are the collapse operators, $A_n$ are the operators that couples the system to environment in $H_I$, and $\gamma_n$ are the corresponding rates.  The density matrix is assumed to be initially in the product state as $\rho_T(0)=\rho(0)\otimes\rho_E(0)$, i.e., the original system and the environment are not correlated at $t=0$. They still remains separable $\rho_T(t)\approx\rho(t)\otimes\rho_E$ during the evolution since the environment does not evolve significantly. The environment is considered to be Markovian, requiring the fast decays of its correlation functions than those of the system.  it is also worthwhile to mention that the evolution of an open quantum system is no longer unitary. Therefore, the density matrix of the quantum system is not a pure state, but a mixed state $\rho=\sum_np_n\ket{\psi_n}\bra{\psi_n}$ instead, with $p_n$ be the classical probability of being in $\ket{\psi_n}$ state.

\subsection{Weak measurement and continuous measurement}
A major difficulty of applying ML algorithms in the quantum regime is caused by measurement, which is usually cost-less in the classical realm. The quantum system is destroyed once measurement happens, being projected to an eigenstate once quantum information is extracted. Measuring a wave function by operator $\hat{A}$ outputs eigenvalues, whose expectation follows $\langle\hat{A}\rangle=\bra{\psi}\hat{A}\ket{\psi}$. It can also be expressed in the language of density matrix as $\langle\hat{A}\rangle=\text{Tr}(\rho\hat{A})$. Aharonov's work~\cite{PhysRevLett.60.1351} proposed its extension, which extracts partial information from the quantum system without destroying it. The weak value $A_w=\bra{\psi_f}\hat{A}\ket{\psi_i}/\langle\psi_f\ket{\psi_i}$ is no longer real eigenvalues of the operator, but exotic values instead or even complex, where $\ket{\psi_i}$ and $\ket{\psi_f}$ are pre/post-selected states. The post-selection operation does not always succeed, and the wave function is discarded once the operation fails. Thus, we couple the quantum system to a pointer for entanglement and measure the pointer projectively for a weak value, which is actually the original framework proposed by Aharonov, instead of the later developed pre/post-selection formalism. To be more specific, a Gaussian pointer $\ket{\Phi}=\int(2\pi\sigma^2)^{-1/4}\exp(-q^2/4\sigma^2)\ket{q}dq$ is coupled to the qubit $\ket{\Psi}=[\cos(\alpha/2),\sin(\alpha/2)]^{\text{T}}$, following the interaction Hamiltonian $H_{int}=g(t)p\otimes\hat{A}$, where $\sigma$ is the standard deviation of the pointer's position, $p$ is its conjugate momentum operator, and $g(t)$ is the coupling strength. A non-correlated initial state $\ket{\Phi(q)}\otimes\ket{\Psi}$ is evolved by the Hamiltonian, entangling as $\cos(\alpha/2)\ket{\Phi(q-a_1)}\otimes\ket{a_1}+\sin(\alpha/2)\ket{\Phi(q-a_2)}\otimes\ket{a_2}$, where $a_i$ and $\ket{a_i}$ are the eigenvalues and eigenstates of the operator $\hat{A}$ to be weakly measured, respectively, if $\int_0^{t_0} g(t)dt=1$. If one aims at performing a weak measurement on the Z direction, i.e., the Pauli-Z operator $\hat{A}=\hat{\sigma_z}$, the measurement outputs of the pointer's position follow the probability distribution 
\begin{equation}
P(q)\approx\frac{1}{\sqrt{2\pi\sigma^2}}\exp\left[-\frac{(q-\cos\alpha)^2}{2\sigma^2}\right],
\end{equation}
shifting a displacement of the expectation $\bra{\Psi}\hat{\sigma_z}\ket{\Psi}=\cos\alpha$. Correspondingly, the wave function of the qubit is slightly perturbed as
\begin{eqnarray}
\label{eq:perturbed}
\ket{\tilde{\Psi}}\propto\frac{1}{(2\pi\sigma^2)^{1/4}}\left\{\cos\frac{\alpha}{2}\exp\left[-\frac{(q_0-1)^2}{4\sigma^2}\right]\ket{0}\right. \nonumber\\
\left.+\sin\frac{\alpha}{2}\exp\left[-\frac{(q_0+1)^2}{4\sigma^2}\right]\ket{1}\right\},
\end{eqnarray}
if the weak value $q_0$ is the measurement feedback of the projective measurement.

Furthermore, quantum information can be continuously extracted from the quantum system, realizing a continuous measurement as the information obtained per measurement goes to zero. In this framework, the total operation time is divided into intervals of timestep $\Delta t$,  so that a weak measurement is performed in each interval. The limit $\Delta t\rightarrow 0$ results in continuous measurement, in which stochastic differential equations govern its dynamics~\cite{gross2018qubit,jacobs2006straightforward}. In Fig.~\ref{fig:pipulse}, we illustrate the dynamics of stochastic Schr\"odinger equations, aiming at flipping a qubit by fixed resonant $\pi$-pulse, with different scales of time interval $\Delta t$.

\section{Numerical experiments}
\label{sec:ne}
\subsection{Physical system and task}

As we introduced in Sec.~\ref{sec:models}, systems under quantum noises are no longer governed by Schr\"odinger equation, but Lindblad master equation instead. For pure dephasing, the diagonal Lindblad operators reads $C_n=\sqrt{\gamma_n}\ket{n}\bra{n}$, giving the master equation
\begin{equation}
\dot{\rho} = -\frac{i}{\hbar}[H(t),\rho(t)] + \Gamma\left[\text{diag}(\rho)-\rho\right],
\end{equation}
if $\gamma_0=\gamma_1$, affecting the coherence by reducing the off-diagnol elements of the density matrix. For relaxation, we consider the energy dissipation from the qubit to the external environment on the $X$ direction, i.e., $C=\sqrt{\gamma}\hat{\sigma_x}$. The non-unitary evolution governed by master equation with Lindblad terms converts the density matrix into mixed state,  in which the classical probability $p_n$ of being in $\ket{\psi_n}$ cannot be retrieved. Hence, the perturbed system after weak measurement cannot be analytically calculated by Eq.~\eqref{eq:perturbed}~\cite{PhysRevLett.124.140504}. Here we extend it to the case of the density matrix. Similarly, a Gaussian pointer of pure state $\rho_p=\ket{\Phi}\bra{\Phi}$ is coupled to a two-level system of mixed state $\rho$ by the interaction Hamiltonian $H_{int}=g\delta(t-t')p\otimes\hat{\sigma_z}$. The collective system is evolved from the initial state $\rho_{ini}=\rho_p\otimes\rho$ to $\rho_{fin}$ after the coupling by
\begin{equation}
\rho_{fin}=\exp(-igp\otimes\hat{\sigma_z})\rho_{ini}\exp(igp\otimes\hat{\sigma_z}),
\end{equation}
shifting the pointer by $\langle\hat{\sigma_z}\rangle=\text{Tr}(\rho\hat{\sigma_z})$ when $g=1$. One retrieves the wave pointer after the coupling by tracing out the qubit. The measurement of the pointer's position projects the pointer to its eigenstate $\ket{q_0}$, where the projection operator of the collective system reads $\hat{P}=\ket{q_0}\bra{q_0}\otimes{I}$. In this way, we have the qubit's density matrix after the weak value feedback of $q_0$ by the projection operator and tracing out the pointer.

After clarifying the calculation of state perturbation in the language of the density matrix, we formulate the specific task to be studied by stream learning. We consider the optimal control of a continuously measured qubit within operation time $T$ by ML algorithm. We aim at flipping it from $|0\rangle$ to $|1\rangle$ by a sequence of pulses on the $X$ direction. Each pulse lasts a small interval of $\Delta t$, being described by the driving Hamiltonian $H=\Omega\hat{\sigma_x}$, followed by a weak measurement on the $Z$ direction. We assume that the measurement process is impulsive, i.e., the coupling and projective measurement on the pointer are instant, being independent of the dynamical evolution. Meanwhile, the control pulses may also be imprecise, including slight detuning $H=\Omega\hat{\sigma_x}+\Delta\hat{\sigma_z}$ and amplitude error $\Omega\rightarrow\Omega+\delta\Omega$. The weak value and the last pulse amplitude are fed to the ML model as streaming data. Accordingly, instant feedback from the model controls the quantum system for the next timestep.

\begin{figure*}[h!]
	\centering
	\includegraphics[width=0.8\linewidth]{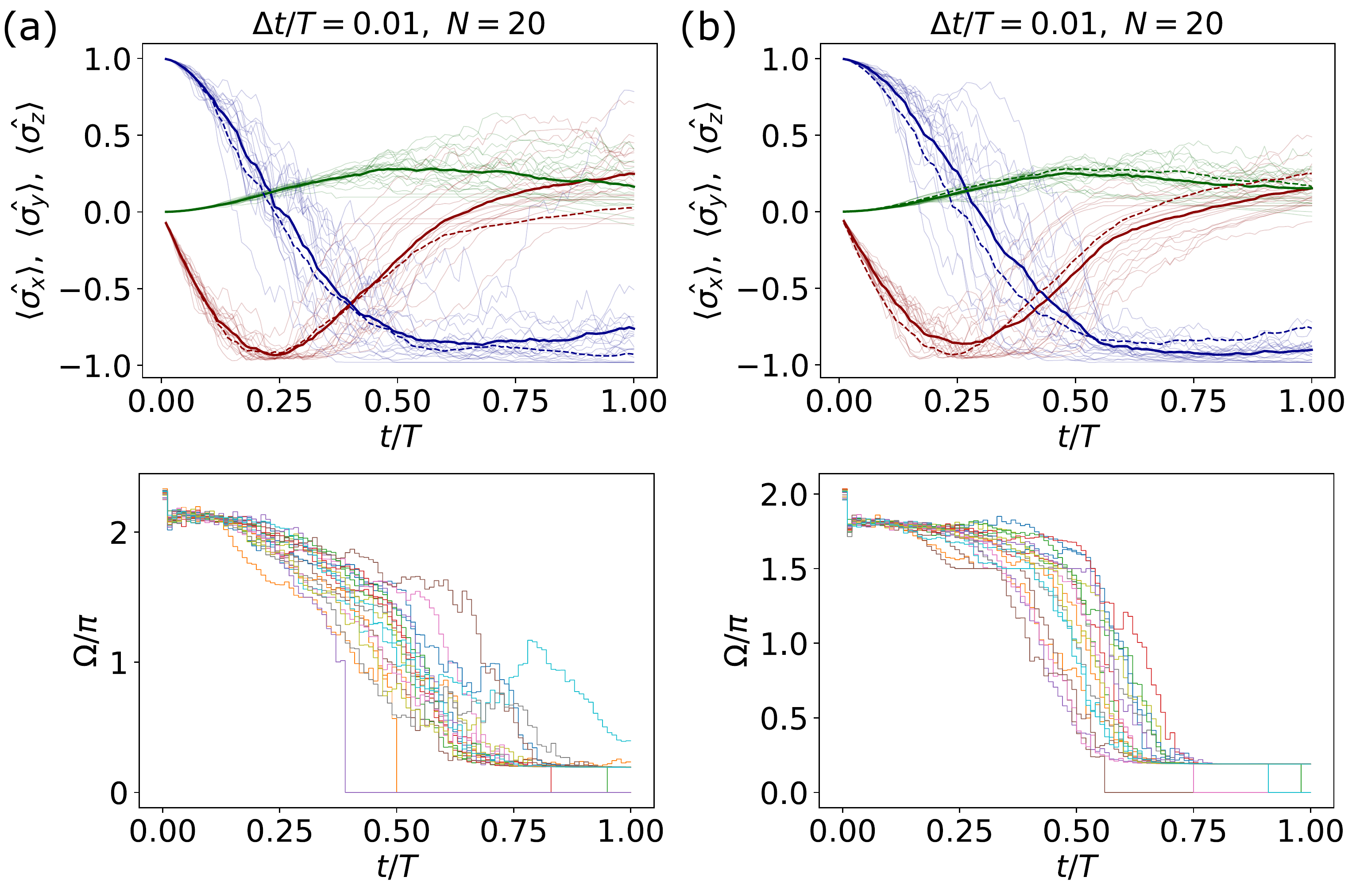}
	\caption{(a) The expectations on $X$ (green), $Y$ (red), and $Z$ (blue) directions of repetitively measured qubits once the agent in Fig.~\ref{fig:threemodels}(c) is employed for driving the qubits in the presence of hybrid detuning errors and quantum noises. The solid curves in Fig.~\ref{fig:threemodels}(c) are plotted in dashed curves as the benchmark. The average fidelity decreases from $0.9636$ to $0.8805$ with the standard deviation of $0.1874$. Imprecise control pulses outputted by the agent as step functions are plotted in different colors. (b) The expectations after training for about 2000 episodes. The solid curves in Fig.~\ref{fig:transfer}(a) are plotted in dashed curves for the benchmark. The agent fits the new environment, retrieving an average fidelity of $0.9513$ with the standard deviation of $0.0353$. Corrected control pulses outputted by the agent as step functions are plotted in different colors.}
	\label{fig:transfer}
\end{figure*}

\subsection{Numerical setup}
We apply the DRL method to our task for the stream learning approach. The environment includes a qubit, which is continuously measured and perturbed for weak values, controlled by the agent's pulses. An artificial neural network (ANN), as the agent, observes a state from the qubit, outputting an action for the control problem, which is trained by a deep learning algorithm for approximating the optimal policy function $\pi(a|s)$. The environment receives the action from the agent, evolving to the next timestep. It also computes the new RL state and the corresponding reward. It is worth noting that the environment in the quantum realm is different from other physical environments. The RL environment requires the numerical simulation if quantum information, e.g., density matrix elements or fidelity, is encoded in the RL state. Instead of a black box in other physical environments, one has to compute the density matrix based on the weak value and the control pulses since elements of the density matrix cannot be obtained from the qubit without destroying it.

In our practice, we set the tunable range of the Rabi frequency (pulse amplitude) as action $\Omega\in[0,3\pi]$ in dimensionless unit, which is renormalized to $\tilde{\Omega}\in[0,1]$ for fitting the neuron. Total operation time $T=1$ is uniformly separated by $n=100$ control pulses. Each pulse drives the qubit for a time interval of $\Delta t=0.01$. For saving the computational resources, the position space of the pointer is within $q\in[-50,50]$, being uniformly separated by $\Delta x=1$, which is large enough for $g=1$. Accordingly, the momentum operator $p$ is constructed by $[q,p]=i\hbar$ with boundary conditions. The density matrix of the collective quantum matrix has a size of $202\times202$. The coarse grained position space leads to weak values $q_0$ of integer number, which is renormalized to $\tilde{q_0}=(q_0+50)/100\in[0,1]$. Thereby, the state is defined as
$
s(t_i)=\{\tilde{\Omega}(t_{i-1}),\tilde{q}_0(t_i),i/n,|\rho_{11}(t_i)|,|\rho_{12}(t_i)|, |\rho_{21}(t_i)|,|\rho_{22}(t_i)|\}$, 
including the last action as renormalized pulse, renormalized weak value, current system time, and elements of the density matrix. The RL state is observed by the agent, an ANN with three fully connected hidden layers of $64$ neurons activated by ReLU, evolving to the next state by the numerical simulation part of the environment, which receives an action from the agent.

\begin{figure*}[h!]
	\centering
	\includegraphics[width=1\linewidth]{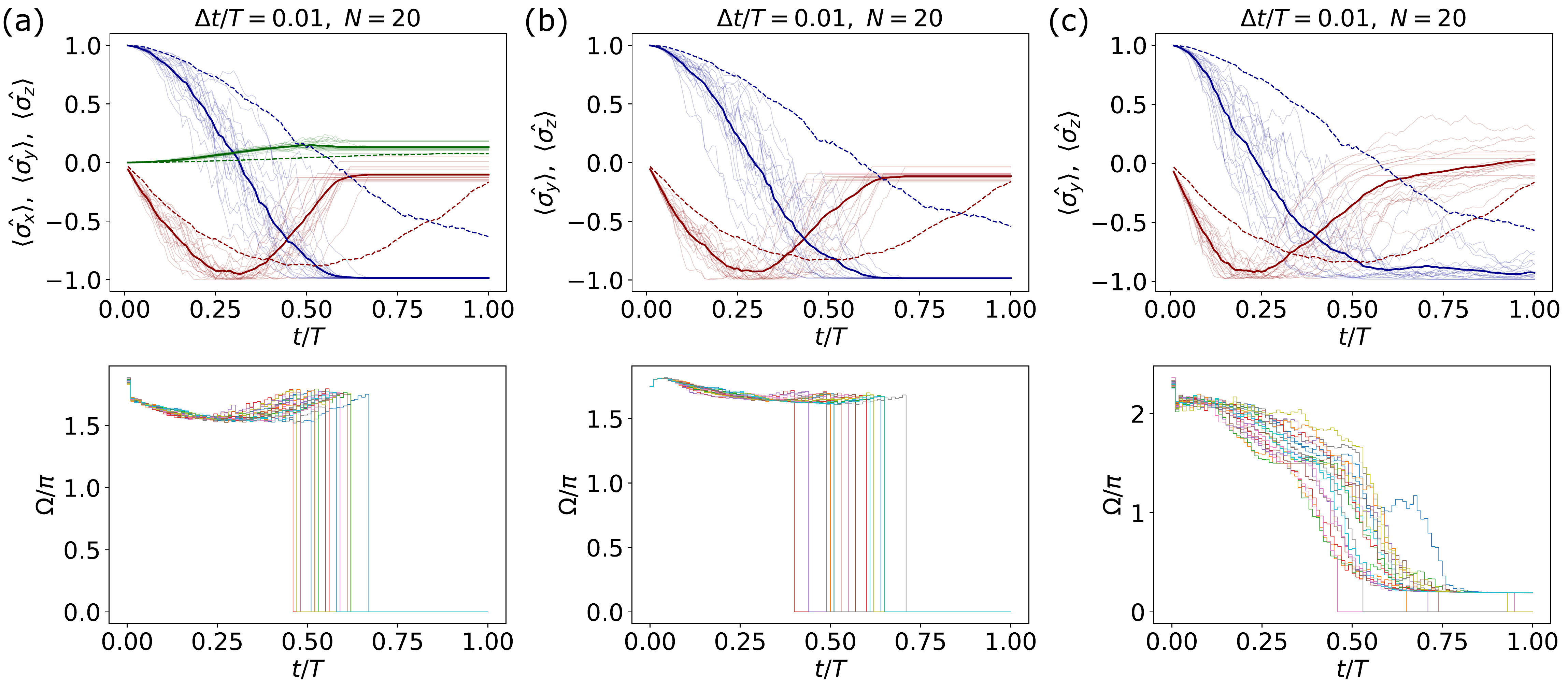}
	\caption{The expectations on $X$ (green), $Y$ (red), and $Z$ (blue) directions of repetitively measured qubits, which are driven by trained DRL agents under (a) detuning, (b) dephasing, and (c) $\hat{\sigma_x}$ relaxation, respectively. The dynamics of each qubit are plotted by shaded curves, being averaged for the solid curve. Fidelities are calculated by $F=\left(\text{Tr}\sqrt{\sqrt{\rho}\ket{1}\bra{1}\sqrt{\rho}}\right)^2$, leading to average fidelities of $0.9922$, $0.9919$, and $0.9636$ for each case, with the standard deviation of $0.0012$, $0.0012$, and $0.0361$. Control pulses provided by the agent as step functions are also plotted in different colors. We set the detuning strength $\Delta=0.05\Omega$, the dephasing rate $\Gamma=0.05$, and the relaxation rate $\gamma=0.05$. Other parameters are the same as those in previous figures, which are listed in the main text. Dashed curves, as baselines,  are derived by averaging the expectations of qubits under $\pi$-pulse control.}
	\label{fig:threemodels}
\end{figure*}

\subsection{Training of the agent and results}
We train three individual models for driving the qubit at the presence of detuning, dephasing, and relaxation on the $X$ direction, respectively. The agents approximate the optimal policy, which maximizes accumulated artificial rewards. We keep the generality in the design of reward functions since we have no specific preference for the shape of the pulses. For the $\ket{0}\rightarrow\ket{1}$ task, we reward the agent by $r(t_i)=|\rho_{22}(t_i)-1|$ per timestep as a negative value, aiming at a fast flipping operation. The agent gets an extra reward of $1000$ if $\rho_{22}$ meets the threshold $|\rho_{22}|>0.99$, and terminates the episode for calculating the total reward early. We also notice that punishment of $100$ if $|\rho_{11}|>0.05$ at the final timestep helps the convergence of the model.

Fig.~\ref{fig:threemodels} shows the high-fidelity closed-loop quantum control at different errors or noises. For relaxation on the $X$ direction, we change the terminal condition by $|\rho_{22}(t_i)|>0.99$ for four neighboring timesteps, which prevents the model from converging on trivial resonant $\pi$-pulses. We used the Proximal Policy Optimization (PPO) method~\cite{schulman2017proximal} to train the agent, with the learning rate being 1e-3 and a batch size of $20$. Other hyperparameters are the default setting by Tensorforce v0.5.3~\cite{tensorforce}. Moreover, we set a random error on the action, characterized by a centered Gaussian distribution with a standard deviation of $0.02$, equivalently implementing the time-varying systematic error in the quantum system. The models give control pulses that are robust against systematic errors. It is worth mentioning that a trade-off usually exists between fidelity and robustness. We obtain the models in Fig.~\ref{fig:threemodels} after about 2000, 3000, and 8000 episodes for controlling the system under detuning, dephasing, and relaxation on the $X$ direction, respectively.

\subsection{Transfer of the agent}
A ML model is online for service after being trained for a particular task, such as flipping a qubit under $\hat{\sigma_x}$ relaxation as Fig.~\ref{fig:threemodels}(c) does. One can evaluate the model by querying the information from the environment to check its validity after it is online. The flipped qubit can go for further tasks, which are independent of the model's duty. In this way, one can also evaluate it by checking the results of additional tasks without querying the environment and the model. Once the performance of a well-trained model deviates from its expectation, one concludes that the qubit in the environment changes, i.e., the quantum errors or noises are shifted to other patterns, if further tasks are reliable or not happening at all. Hence, we want another model for precise control in the new environment. Instead of discarding the current model and training a new one, which is apparently inefficient, we transfer the agent to a new environment for the same task, exploring its capability to fit the new environment without much effort.

We test the proposal by starting from the trained agent in Fig.~\ref{fig:threemodels}(c). By directly evaluating the agent in a new environment in the presence of detuning $\Delta=0.1\Omega$, dephasing rate $\Gamma=0.05$, and $\hat{\sigma_x}$ relaxation rate $\gamma=0.05$, the average final state deviates significantly from the previous result [c.f. Fig.~\ref{fig:transfer}(a)], resulting in a fidelity decrease as well. Furthermore, we train the agent with the same setting for about $2000$ episodes. By additional $20\%$ of the total episodes, the agent retrieves its performance before the environment shift happens [c.f. Fig.~\ref{fig:transfer}(b)].

\section{Discussion and Outlook}
Based on the numerical experiment in Sec.~\ref{sec:ne}, we reckon that DRL can be employed as a stream learning approach to closed-loop quantum control. The fidelity can be further improved by fine-tuning in another training environment, with different thresholds and the design of reward function. Interestingly, we found out that the optimized policy from the agent is interpretable to some extent, as Fig.~\ref{fig:threemodels}(c) and Fig.~\ref{fig:transfer} shown. The maximal tunable Rabi frequency is $3\pi$, which is $1.5$ times the $\pi$-pulse for an operation time of $T=1$. The agent drives the qubit with a relatively high frequency for reaching a large $\rho_{22}$ sooner. It is understandable since continuous measurement can be described in the language of superoperators, affecting the dynamics like quantum noises, which effectively can be suppressed by reduced operation time. Later, the pulse strength decreases significantly once $\rho_{22}$ is large enough, converging to a small constant value for more precise operations. Accordingly, it is the weak measurement predominant in the state evolution instead of the control pulse. It is also similar to the quantum Zeno effect that locks a wave function on its eigenstate by repeating projective measurements. 

Now we further discuss this topic after analyzing the results above. In Sec.~\ref{sec:ne}, we explained that the RL environment consists of a qubit and a numerical simulation part. The qubit can be physical, e.g., constructed in superconducting circuits, trapped ions, photonics, etc., or simulated by classical computers as we performed in numerical experiments. Here we emphasize again that the numerical simulation for calculating the qubit dynamics is compulsory if we include quantum information $\rho_{ii}$ or fidelity in the RL state and reward. Although we can perform the weak measurement, extracting partial quantum information and converting it to weak value $q_0$ without destroying the quantum state, it is still impossible to retrieve the total information of the density matrix by a single shot of measurement. We cannot treat the qubit as a black box, as we usually do in other classical scenarios, where the observation of the RL state is instant and cost-less. By contrast, we have to calculate the qubit dynamics based on the actions and feedback, deducing $\rho_{ii}$ without operating on the qubit. It becomes a setback when one performs stream learning in the quantum realm since simulating the quantum dynamics is time-consuming, e.g., about $7$ seconds for an episode in our numerical experiment. However, stream learning for real quantum devices requires the simulation speedup of about $10^5$ times (compared to the T1 time of state-of-the-art superconducting qubit). A possible solution is to train another ANN to mimic the dynamics of the quantum system, with available information as input, outputting the quantum information to be deduced without measurements. The training of such ANN needs plenty of training data and adequate training methods, which goes beyond the scope of this work.

Another method to avoid the black-box problem is to exclude the quantum information in the RL state. The RL state may contain the weak value $q_0$ and other classical information such as last action, the system time, etc.  It should be reminded that the threshold criteria for early termination are no longer available in this paradigm. Thus, the training environment only rewards the agent by a constant at the end of each episode once a projective measurement on the target state succeeds. However, the agent struggles to learn the precise control since the signal-to-noise ratio of $q_0$ is too low. The reward criteria also need a large ensemble (batch size) to evaluate the fidelities of final  quantum states. In this way, the problem becomes more challenging, which can be applied for evaluating RL algorithms.

\section*{Acknowledgments}
This work is partially supported from NSFC (12075145), STCSM (2019SHZDZX01-ZX04 and 20DZ2290900), SMAMR (2021-40), Program for Eastern Scholar, QMiCS (820505) and OpenSuperQ (820363) of the EU Flagship on Quantum Technologies, EU FET Open Grant Quromorphic (828826) and EPIQUS (899368), Spanish Government PGC2018-095113-BI00 (MCIU/AEI/FEDER, UE), and Basque Government IT986-16. X.C. acknowledges the Ram\'on y Cajal program (RYC-2017-22482).

\bibliographystyle{ieeetran}
\bibliography{SLQ}

\end{document}